\documentclass[12pt]{iopart}
\usepackage{graphicx}%
\usepackage{amssymb}
\usepackage{mathrsfs}%

\begin{document}

\title[Shot noise enhancement]{Shot Noise Enhancement from Non-equilibrium
  Plasmons in Luttinger-Liquid Junctions}%

\author{Jaeuk U Kim\dag, Jari M Kinaret\ddag~and Mahn-Soo Choi\S}%
\address{\dag Department of Physics, G\"oteborg University,
SE-41296 Gothenburg, Sweden}
\address{\ddag Department of Applied Physics, Chalmers University of
  Technology, SE-41296 Gothenburg, Sweden}
\address{\S Department of Physics, Korea University, Seoul 136-701,
  Korea}%
\ead{choims@korea.ac.kr}%

\begin{abstract}
We consider a quantum wire double junction system with each wire segment
described by a spinless Luttinger model, and study theoretically shot
noise in this system in the sequential tunneling regime.
We find that the non-equilibrium
plasmonic excitations in the central wire segment give rise to
qualitatively different behavior compared to the case with equilibrium
plasmons.  In particular, shot noise is greatly enhanced by
them, and exceeds the Poisson limit.
We show that the enhancement can be explained by the emergence
of several current-carrying processes, and that the effect disappears
if the channels effectively collapse to one due to, {\em e.g.}, fast
plasmon relaxation processes.
\end{abstract}

\pacs{71.10.Pm, 72.70.+m, 73.23.Hk, 73.63.-b}%
\maketitle%

\let\eps=\epsilon%
\let\veps=\varepsilon%
\let\what=\widehat%
\let\tol=\leftarrow
\newcommand\avg[1]{\left\langle\textstyle#1\right\rangle}%
\newcommand\set[1]{\left\{\textstyle#1\right\}}%
\newcommand\ket[1]{\left|\textstyle#1\right\rangle}%
\newcommand\bra[1]{\left\langle\textstyle#1\right|}%
\newcommand\braket[1]{\left\langle\textstyle#1\right\rangle}%
\newcommand\hatI{\hat{I}}%
\newcommand\half{\frac{1}{2}}%
\newcommand\whatbf[1]{\what{\mathbf{\textstyle#1}}} %
\newcommand\hatbf[1]{\hat{\mathbf{\textstyle#1}}} %

\section{Introduction}
\label{noise-llj::introduction}

Shot noise in electronic devices occurs due to the discrete nature of
the transported charge, and manifests itself clearly in non-equilibrium
situations~\cite{Schottcky18a,Hull25a}.  Unlike thermal (equilibrium)
noise, shot noise provides additional information of the system, which
is not accessible in an average current
measurement~\cite{Davies92a,Wilkins92a,Hershfield93a}.  For this reason,
shot noise has been a subject of a great number of theoretical and
experimental works in recent years~\cite{Blanter00a}.

One of the fundamental issues of current interest has been the effect of
coherence on the shot noise~\cite{Oberholzer02a}.  Somewhat contrary to
intuition, it has been shown that in most cases where comparable
theories are available, the semiclassical and quantum mechanical
descriptions give the same result; it seems that often the shot noise is not
sensitive to the quantum
coherence~\cite{Blanter00a}. 
Another important issue related to the shot noise is electron-electron
correlation effects, which is the subject of the present work.  Popular
examples are the shot noise in a single-electron transistor (SET) in the
Coulomb blockade (CB) regime~\cite{Hershfield93a,Korotkov94a} and in
superconducting
tunnel junctions~\cite{Deblock03a,ChoiMS03a,ChoiMS01c}. 
The system of our interest in this work is a one-dimensional (1D)
interacting electron gas described by the Luttinger liquid (LL)
model~\cite{Haldane81b}.  There has been a great deal of experimental
effort to demonstrate that certain low-dimensional systems (e.g., carbon
nanotubes) behave as Luttinger
liquids~\cite{bockrath99a,yao99a,postma01a}.
Shot noise in Luttinger liquid has also been the subject of many
recent works. For example, the nonequilibrium shot noise of the
edge states was used to measure the fractional charge of the
quasiparticles of the fractional quantum Hall
states~\cite{Kane94a,Saminadayar97a,Picciotto97a,Comforti02a}.

Very recently, Braggio \etal \cite{Braggio03a} investigated the
shot noise in a double junction system embedded in a Luttinger
liquid. They found some features due to the interplay between the
CB, the LL behavior, and size quantization.  Apart from the
peculiar \emph{quantitative} behavior of the shot noise at higher
bias voltages, its qualitative behavior as a function of bias
voltage was similar to that in conventional SET devices.  They
assumed equilibrium distribution of charge-density excitations (in
the following called plasmons) in all electrodes. 
On the other hand, in a recent work by some of the present authors
\cite{KimJ03a}, plasmons in the short central electrode in the
absence of environment coupling were found to exhibit a highly
non-equilibrium distribution. 
Since shot noise is sensitive to non-equilibrium fluctuations in the
system, one can expect that the non-equilibrium plasmons
may affect the shot noise significantly. 
In this work, we show that the non-equilibrium plasmons lead to
\emph{qualitatively} different behavior of the shot noise even at
relatively low bias voltages.  We also show that in the limit of fast
plasmon relaxation, when the plasmon distribution reaches equilibrium
between successive tunneling events, the results by Braggio
and co-workers are reproduced. 

\section{Model}
\label{noise-llj::model}

We consider a LL/LL/LL double junction with each LL representing a
segment of quantum wire.  We refer to the three segments the left ($L$) and
right ($R$) leads and the dot ($D$), respectively.
The two leads are adiabatically connected to reservoirs (contacts) and
the dot (the central short wire segment) is coupled to the two leads by
tunnel junctions.
The L and R leads are described by semi-infinite (spinless) Luttinger
models while the dot is described by a finite (spinless) Luttinger
model~\cite{Braggio03a,KimJ03a}.
The low-energy transport properties through the system are described by
the master equation~\cite{KimJ03a,Braggio00a} for the probability
$P(N,\set{n},t)$ to find $N$ (excess) electrons and $n_m$ plasmons at
mode $m$ on the dot at time $t$.
The eigenstates of an isolated dot are denoted by $\ket{N,\set{n}}$. The
finite length $L_D$ of the dot leads to discrete energy spectrum of
plasmons $\set{n}$ and the zero-mode energy associated with $N$. The
equidistant plasmon spectrum is characterized by the energy spacing
\begin{math}
\veps_p \equiv {\pi\hbar v_F}/{L_Dg}
\end{math},
where $g$ the Luttinger parameter and $v_F$ the Fermi velocity.  The
zero-mode energy, characterized by the energy scale $E_C=\veps_p/g$,
accounts for the charging effects on the dot.
Electrons are transferred between different wire segments by
tunneling, and the transition rates $\Gamma_{L/R}(N\pm
1,\set{n'};N,\set{n})$ between states $\ket{N,\set{n}}$ due to
tunneling across the two junctions are obtained using Fermi golden
rule~\cite{KimJ03a}. Following the notations of Ref.
\cite{KimJ03a}, we consider that the LL-parameter $g$ is the same
for each wire segment, the capacitances across L/R tunneling
junctions are taken to be the same ($C_L=C_R$), but the tunneling
amplitudes may differ and their ratio is denoted by
$R=|t_L|^2/|t_R|^2$. The explicit form of the transition rates for
a wire with a single branch of spinless fermions at applied
voltage $V=V_L-V_R$ is given by
\begin{equation}
\begin{array}{l}
\Gamma_{L/R}(N,\{n\} \rightarrow N^\prime,\{n^\prime\}) =
{2\pi|t_{L/R}|^2}/ {\hbar} \\
\times
\gamma(E_D(N^\prime,\{n^\prime\})-E_D(N,\{n\})\mp (N^\prime- N) eV_{L/R})  \\
  \times |\langle N^\prime,\{n^\prime\}|
\psi^\dagger_D\delta_{N^\prime,N+1}+\psi_D\delta_{N^\prime,N-1}| N,\{n\}
\rangle|^2,
\end{array}
\label{eq:Gammarate}
\end{equation}
where the contribution of the leads is
\begin{equation}
\begin{array}{rl}
\gamma(\epsilon)=
& \begin{displaystyle}
\frac{1}{2\pi\hbar}\frac{1}{\pi v_F}\left(\frac{2\pi \Lambda_g}{\hbar v_F\beta}
\right)^\alpha\left|\Gamma(\frac{\alpha+1}{2}+i\frac{\beta\epsilon}{2\pi})\right|^2
\frac{e^{-\beta\epsilon/2}}{\Gamma(\alpha+1)}
\end{displaystyle}
\end{array}
\label{eq:gammalead}
\end{equation}
and the contribution of overlap matrix elements in the central segment is
\begin{equation}
\begin{array}{l}
|\langle \{n^\prime\}|\psi^\dagger_{D}|\{n\}\rangle|^2
=\frac{1}{L_D}\left(\frac{\pi \Lambda}{L_D}\right)^{\alpha} \times\\
\begin{displaystyle}
 \prod_{m=1}^{\infty} \left(\frac{1}{mg}\right)^{|n_m^\prime-n_m|}
\frac{n_m^{(<)} !}{n_m^{(>)} !}
\left[ L^{|n_m^\prime-n_m|}_{n_m^{(<)}}\left(\frac{1}{mg}\right)\right]^2
\end{displaystyle}
\end{array}
\label{eq:Laguerre}
\end{equation}
where $n_m^{(<)}=\min(n_m^\prime,n_m)$ and $n_m^{(>)} = \max(n_m^\prime,n_m)$.
Here $L^n_m(z)$ is a Laguerre polynomial, $\Lambda$ a short-distance
cutoff, and $\alpha = (g^{-1}-1)$ the appropriate exponent for
tunneling into the end of a Luttinger liquid. Note that while the lead
contribution depends only on the energies $E_D$ of the involved states
of the dot, the contribution arising from the dot depends on the detailed
nature of the participating states. Temperature enters the rates through
$\beta = 1/(k_BT)$; since we are interested in shot noise rather than thermal
noise, we from now on
set $k_BT$ to a value far below any relevant energy scale
in the dot.

The plasmons in the two leads, being in contact with reservoirs
with many low-energy excitations, are
assumed to be in equilibria separately.  However, the plasmons in the
dot are driven far from the equilibrium distribution~\cite{KimJ03a} by
tunneling electrons. This deviation is already significant at relatively
low bias voltages $2\veps_p \leq eV \leq 2E_C$. Contrast to the
equilibrium plasmons \cite{Braggio03a} that produce qualitatively
similar shot noise as in usual SET~\cite{Hershfield93a}, the
non-equilibrium distribution strongly affects the SN of the
system.

The coupling of the system to the environment causes even the plasmon
distribution in the dot to relax towards an equilibrium. The precise
form of the relaxation rate, $\Gamma_p$, depends on the specific relaxation
mechanism, but the physical properties of our concern are not sensitive
to such details.  We use a phenomenological model, obeying detailed
balance, with
\begin{equation}
\label{noise-llj::eq:Gp} \centering\Gamma_p(\set{n'},\set{n}) =
\gamma_p \frac{W_p/\veps_p}{e^{\beta W_p} - 1}
\end{equation}
where
\begin{math}
W_p = \veps_p\sum_m m(n_m'-n_m) \end{math} 
is the energy difference of the two many-body states with $\{n'\}$ and
$\{n\}=(n_1,n_2,\cdots,n_m,\cdots)$ plasmon occupations, and
$\beta=1/k_BT$ is the inverse temperature. %
While the plasmon relaxation rate in nanoscale structures is difficult
to estimate, recent computer simulations on carbon nanotubes indicate
that plasmon excitations in them have life times of the order of a
picosecond, much longer than those in three-dimensional structures
\cite{Tomanek_pc}. The total transition rate is hence given by a sum of
rates associated with tunneling and plasmon relaxation,
\begin{math}
\Gamma(N',\set{n'};N,\set{n}) =
\sum_{\ell=L,R}\Gamma_\ell(N',\set{n'};N,\set{n}) + \delta_{N',N}~
\Gamma_p(\set{n'},\set{n})
\end{math}
where $\ell$ labels the tunnel junctions. The only phenomenological
parameter is the time scale of the relaxation processes in the dot,
$\gamma_p$; all other quantities are determined by the geometry of the
system and the properties of the tunnel junctions.

The master equation approach neglects phase coherence of electrons
tunneling across the two junctions (sequential tunneling regime).  The
effects of the coherence on the SN have been hotly debated in
recent years.  However, for many mesoscopic systems (including
low-dimensional interacting systems), the master equation approach
yields the same results as the fully quantum mechanical description of
the SN~\cite{Blanter00a}.  Therefore, we expect that our master
equation approach yields qualitatively correct results for the system at
hand.

We introduce a matrix notation for the transition rates $\Gamma$ with
matrix elements defined by
\begin{equation}
\label{noise-llj::eq:G11}
\left[\widehat\Gamma_\ell^\pm(N)\right]_{\{n^\prime\},\{n\}} =
\Gamma_\ell(N\pm 1,\set{n'};N,\set{n}) \,,
\end{equation}
\begin{equation}
\label{noise-llj::eq:G12}
\left[\widehat\Gamma_\ell^0(N)\right]_{\{n'\},\{n\}} \\\mbox{}%
= \delta_{\{n'\},\{n\}} \sum_{\{n''\}} \left[\widehat\Gamma_\ell^+(N)
  + \widehat\Gamma_\ell^-(N)\right]_{\{n''\},\{n\}} \,,
\end{equation}
and
\begin{equation}
\label{noise-llj::eq:G13}
\left[\widehat\Gamma_p(N)\right]_{\{n'\},\{n\}}
= - \Gamma_p(\set{n'},\set{n}) \\\mbox{}%
+ \delta_{\{n'\},\{n\}}\sum_{\{n''\}} \Gamma_p(\set{n''},\set{n}) \,.
\end{equation}
Using this notation, the master equation is
\begin{equation}
\label{noise-llj::eq:ME2} \frac{d}{dt}\ket{P(t)} =
-\widehat\Gamma\ket{P(t)}
\end{equation}
with
\begin{math}
\widehat\Gamma =
\widehat\Gamma_p + \sum_{\ell=L,R}
\left(\widehat\Gamma_\ell^0 -
  \widehat\Gamma_\ell^+ - \widehat\Gamma_\ell^-\right)
\end{math},
where $\ket{P(t)}$ is the column vector (not to be confused with the
``ket'' in quantum mechanics) with elements given by
\begin{math}
\braket{N,\set{n}|P(t)} \equiv P(N,\set{n},t)
\end{math}.

\section{Shot Noise}
\label{noise-llj::shotnoise}

We investigate the shot noise power defined by
\begin{equation}
\label{noise-llj::eq:SN1}
S(\omega) = 2\int_{-\infty}^\infty{d\tau}\;
e^{+i\omega\tau}\left[
  \avg{I(t+\tau)I(t)} - \avg{I(t)}^2
\right]
\end{equation}
in the steady state ($t\to\infty$).  In terms of the \emph{tunneling}
current $I_{L/R}$ across the junction $L/R$, the \emph{total} current
$I$ is conveniently written as
\begin{math}
I = \sum_{\ell=L,R}(C/C_\ell)I_\ell
\end{math}
where $C^{-1} = C_L^{-1} + C_R^{-1}$.  The correlation functions
$K(\tau)=\avg{I_\ell(t+\tau)I_{\ell'}(t)}$ ($t\to\infty$) can be deduced
from the master equation (\ref{noise-llj::eq:ME2}); in the matrix
notation~\cite{Hershfield93a,Korotkov94a} we have
\begin{eqnarray}
K(\tau) &=& e^2\sum_{N,\set{n}} \bra{N,\set{n}} \Bigg[
\Theta(+\tau) \hatI_\ell\exp(-\widehat\Gamma\tau)\hatI_{\ell'}
\nonumber
\\\mbox{}%
&&+ \Theta(-\tau) \hatI_{\ell'}\exp(+\widehat\Gamma\tau)\hatI_\ell
+ \delta(\tau)\delta_{\ell\ell'} \left(\widehat\Gamma_\ell^+ +
  \widehat\Gamma_\ell^-\right) \Bigg] \ket{P(\infty)} \,,
\label{eq:K}
\end{eqnarray}
where $\Theta(x)$ is the unit step function, $\ket{P(\infty)}$ is the
steady-state solution to the master equation (\ref{noise-llj::eq:ME2}),
and $\widehat{I}_{L/R}$ are tunneling current matrices
\begin{math}
\widehat{I}_{L/R} = \mp e\left(\widehat\Gamma_{L/R}^+ -
  \widehat\Gamma_{L/R}^-\right)
\end{math}.

\begin{figure}[b]
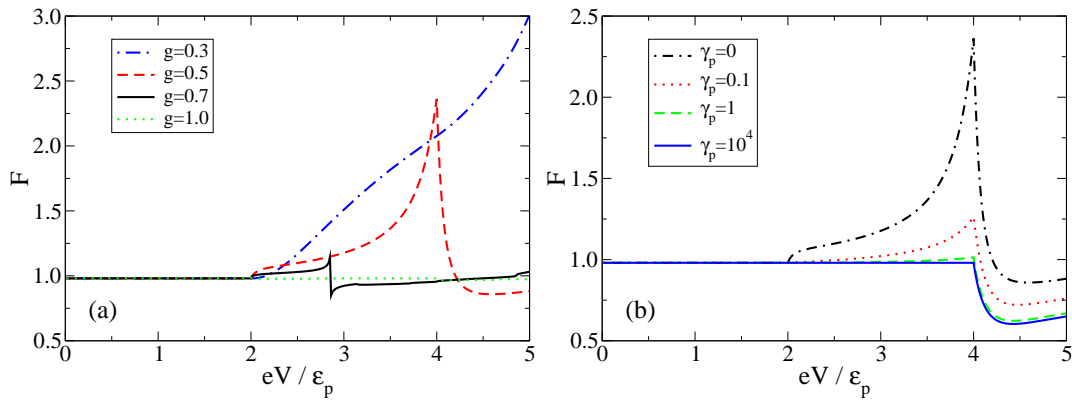

\centering%
\includegraphics*[width=7cm]{Fig1a}
\includegraphics*[width=7cm]{Fig1b}
\caption{Fano factor $F\equiv S(0)/2e\avg{I}$ as a function of
  bias voltage (a) for $g=0.3, 0.5, 0.7, 1$, with no plasmon relaxation
  on the dot, and (b) for $g=0.5$ with different plasmon relaxation
  rates ($\gamma_p$).  Here $R=100$ (highly asymmetric junctions), and
  $T=0$.}
\label{noise-llj::fig:1}
\end{figure}

For bias voltages and temperatures when only a few levels are
involved in the transport, the Fano factor can be evaluated
analytically, but the general case requires numerical integration
of equation (\ref{noise-llj::eq:ME2}).  Figure
\ref{noise-llj::fig:1} shows a typical behavior of the Fano factor
as a function of bias voltage for different values of the
interaction parameter $g$. The dips at $eV=2E_C=2\veps_p/g$ for
$g=0.7$ and $g=0.5$ in figure \ref{noise-llj::fig:1} (a) are
ascribed to the charging effect as typically seen in SET
devices~\cite{Hershfield93a,Korotkov94a}. This dip structure was
also observed by Braggio \etal \cite{Braggio03a}.

A remarkable feature that contrasts our model with a conventional
SET~\cite{Hershfield93a,Korotkov94a}, and the model in
Ref.~\cite{Braggio03a}, is the enhancement of the shot noise in the bias
range $2\veps_p< eV < 2E_C$; see figure \ref{noise-llj::fig:1} (a).  As
the external bias voltage increases over the threshold value $2\veps_p$,
it provides an electron with enough energy to tunnel across the double
junction through the plasmon excitation level of mode $m=1$, and hence
the current sharply increases~\cite{KimJ03a,Braggio00a}.  Unlike a usual
resonance channel, however, this additional channel is very noisy in the
sense that the Fano factor exceeds the conventional
Poisson limit ($F=1$). 
Therefore, the statistics of the tunneling events across the double
junction through the plasmon excitations are highly non-Poissonian.

We emphasize that the SN enhancement is due to the non-equilibrium
distribution of the plasmons on the dot.  To see this, we turn on the
plasmon relaxation, $\gamma_p\neq 0$.
For sufficiently large relaxation rates, any plasmon excitations
induced by electron tunneling relax to an equilibrium before a
subsequent tunneling event.  Therefore, as the plasmon relaxation
rate increases, the Fano factor should recover the characteristic
of the conventional SET or the model of Braggio \etal
\cite{Braggio03a}. 
This is clearly seen in figure \ref{noise-llj::fig:1} (b), where the Fano
factor is shown as a function of $eV$ for relative plasmon relaxation
rates $\gamma_p/\Gamma_0=0,0.1,1,10^4$, where
$\Gamma_0=|t|^2/\hbar^2v_FL_D$ with $|t|^{-2} =
|t_L|^{-2}+|t_R|^{-2}$.

A simple explanation of the enhanced noise may be obtained by
considering a situation in which only three states of the dot are
energetically allowed. We denote the states $|0\rangle$,
$|1A\rangle$ and $|1B\rangle$ where the integer denotes the (excess)
number of charges, and $A$ and $B$ label different states with the
same particle number\footnote{In the case studied above, this three-state
model is reasonably accurate for $2\varepsilon_p\lesssim eV\ll 2E_C$, when
$A$ corresponds to the state with no plasmon excitations and $B$
to the state with one plasmon at the mode $m=1$ ($n_1=1$).}.
The transition rates between these states,
caused by tunneling event into and out of the dot, are given by
$\Gamma_\alpha^s$ where $s = \pm$ indicates if the transition is
from charge state 0 to charge state 1 or vice versa, and the label
$\alpha = A,B$ tells which of the states $|1A\rangle$ and
$|1B\rangle$ is involved in the transition. Hence, there are two
current carrying processes involving transitions $|0\rangle
\rightarrow |1\alpha\rangle \rightarrow |0\rangle$, and, in a time
sequence, the two processes alternate randomly. This extra degree of
freedom results in additional noise, and is reflected in the Fano
factor (which can be calculated as described above)
\begin{equation}
F = 1 + \frac{\Gamma_A^+\Gamma_B^+(\Gamma_A^--\Gamma_B^-)^2 -
\Gamma_A^-\Gamma_B^-(\Gamma_A^-\Gamma_B^+ + \Gamma_A^+\Gamma_B^-)}
{(\Gamma_A^+\Gamma_B^- + \Gamma_A^-\Gamma_B^+ +
\Gamma_A^-\Gamma_B^-)^2} \label{eq:F2}
\end{equation}
that can exceed unity if the two processes correspond to very
different current levels; in the case studied above this happens
when the interaction is sufficiently strong. If there is a fast
internal relaxation
process that allows the state $|1B\rangle$ to decay to $|1A\rangle$,
the two channels effectively collapse into one, and the Fano factor
is given by
\begin{equation}
F = \frac{[\Gamma_A^-]^2 + [\Gamma_A^+ + \Gamma_B^+]^2} {[\Gamma_A^-
+ \Gamma_A^+ + \Gamma_B^+]^2} \in [\frac{1}{2},1] \label{eq:F1}
\end{equation}
which is the result for a single channel system with transition
rates $\Gamma^- = \Gamma_A^-$ and $\Gamma^+ = \Gamma_A^+ +
\Gamma_B^+$. Hence, the presence of charge-conserving internal
relaxation processes reduces the effective number of transport
processes, and the noise associated with the random alternation
between them.

\begin{figure}
\includegraphics[width=6.7cm]{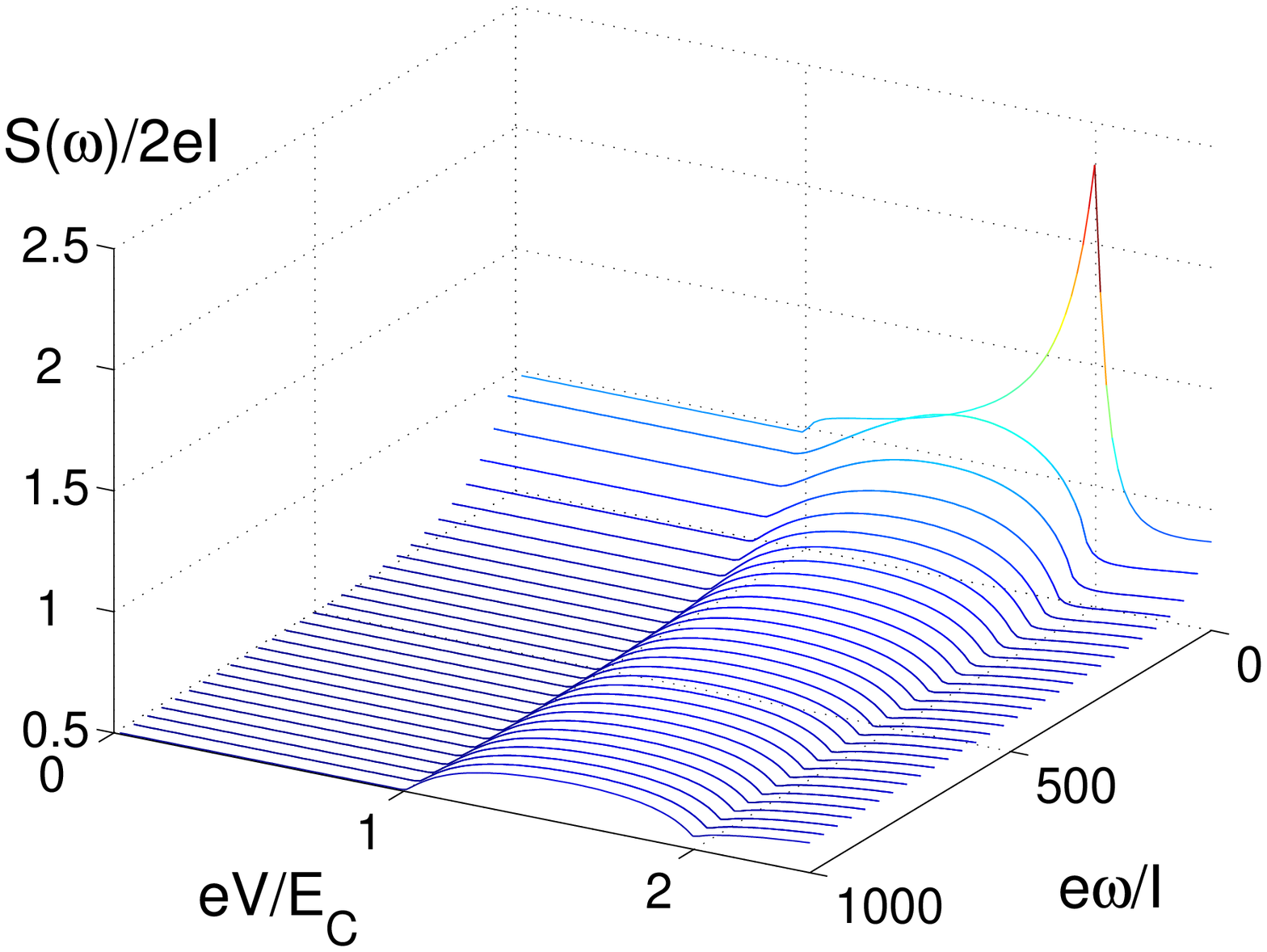} (a)
\includegraphics[width=6.7cm]{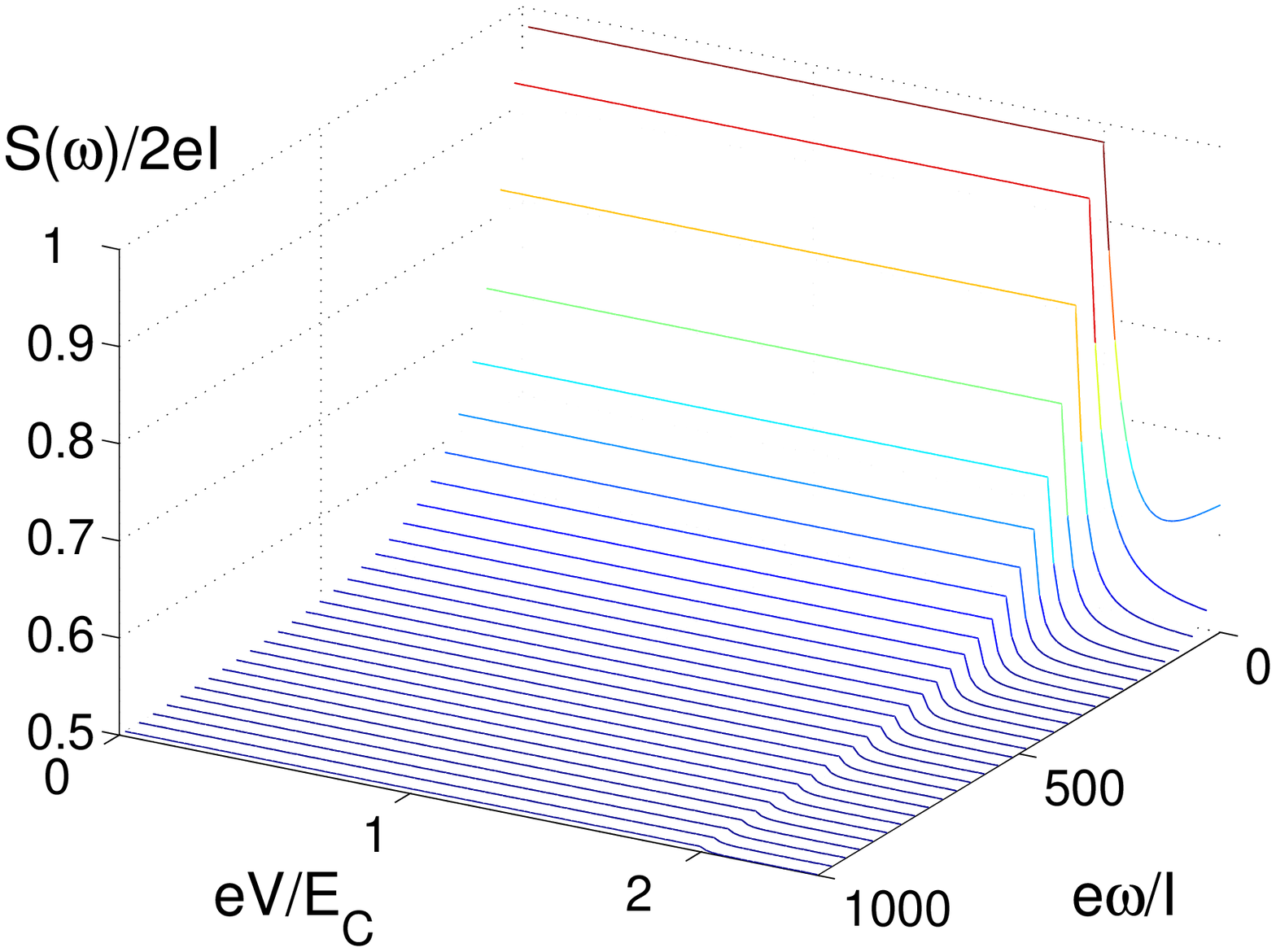} (b) \\
\includegraphics[width=6.7cm]{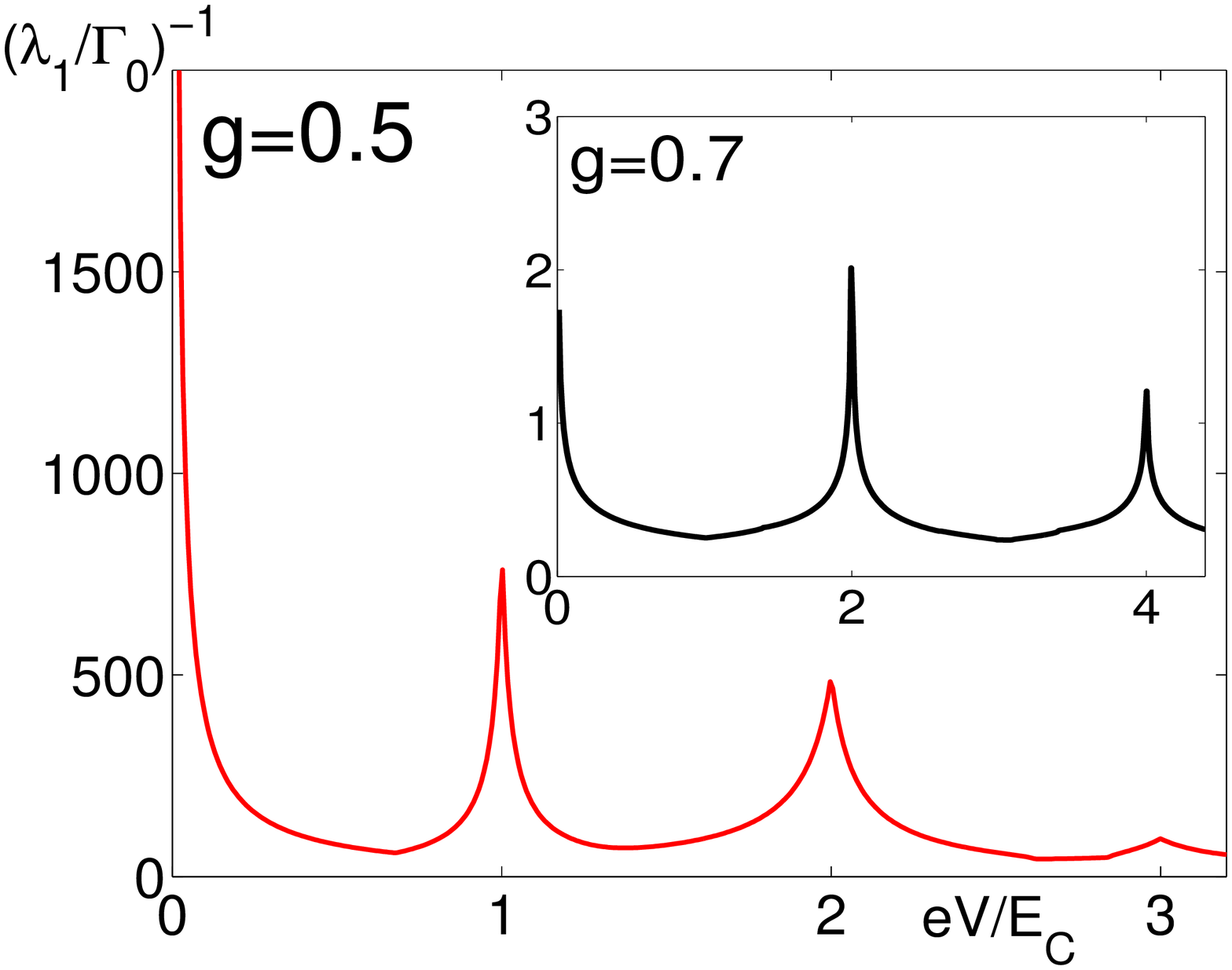} (c) \\
\caption{Finite frequency shot noise $S(\omega)/2eI$ as a function of
the bias $eV/E_C$ and frequency $e\omega/I$ for $R=100$ and $g=0.5$,
at $N_G=1/2$ ($T=0$), (a) with no plasmon relaxation
($\gamma_p=0$) and (b) with fast plasmon relaxation
($\gamma_p=10^4$). In (c), the longest relaxation time with
$\gamma_p = 0$ as a function of the applied voltage for strong
and weak interactions ($g = 0.5$ and $g=0.7$, respectively). The
relaxation time is given in units of
$\Gamma_0=|t|^2/\hbar^2v_FL_D$ with $|t|^{-2} =
|t_L|^{-2}+|t_R|^{-2}$.  }
\label{fig:FWeV}
\end{figure}

The shot noise enhancement persists to quite high frequencies as
shown in figure \ref{fig:FWeV}, where the shot noise power
$S(\omega)/(2eI)$ is shown as a function of the bias and frequency
for a strong interaction ($g=1/2$) and asymmetric tunnel barriers ($R
= 100$) without plasmon relaxation (a) and with fast plasmon
relaxations (b).

The frequency dependence of noise reveals two additional
consequences of non-equilibrium plasmons. Firstly, the asymptotic
high-frequency value of $S(\omega)$ is increased by the
non-equilibrium distribution, and secondly, the characteristic
frequency scale of the noise is seen to depend quite sensitively on
the applied voltage.

In the high frequency limit, $\omega \rightarrow \infty$, the
correlation effects are lost in the noise power except the
$\delta(\tau)$-term in (\ref{eq:K}) that reflects the Pauli
exclusion~\cite{Korotkov94a}, and the asymptotic value of the
noise spectrum reduces to
\begin{equation} \label{eq:SN_infty}
S(i\omega\rightarrow\infty) = 2e\left(\frac{C_R^2A_L}{(C_L+C_R)^2}+
\frac{C_L^2A_R}{(C_L+C_R)^2}\right) =\frac{e}{2}(A_L+A_R),
\end{equation}
where $A_{L/R}$ in the steady-state is
given by
\begin{equation}
\label{Noise-llj-long::eq:A}
A_{\ell}=e \sum_{N,\{n\}}
\langle N,\{n\}|\whatbf{\Gamma}^+_{\ell}
+\whatbf{\Gamma}^-_{\ell}|P(\infty)\rangle \,.
\end{equation}
The quantity $A_{\ell}$ counts the total number of tunneling events
across the junction $\ell$, without regard to direction. In the
absence of tunneling events against the voltage, which is the case
e.g. in the limit of fast plasmon relaxation and low temperatures,
the limiting value of $S(\omega)$ is simply proportional to the
current. In the non-equilibrium case, however, some plasmon modes
with sufficiently high energy are occupied so that tunneling against
the voltage is not negligible, which results in the enhanced
high-frequency noise above the customary limiting value
$S(\omega\rightarrow\infty) = eI$.

The frequency dependence of $S(\omega)$ comes from terms of the
form $\lambda_j/(\omega^2 + \lambda_j^2)$ (see \Eref{eq:K}), where
$\lambda_j$ are the non-zero eigenvalues of the transition matrix
$\Gamma$, that is, the characteristic relaxation rates of the
system. The longest relaxation time, shown in figure
\ref{fig:FWeV}(c) for $g=0.5$ and $g=0.7$, exhibits maxima at
threshold voltages when more states (either charge states or
additional plasmon excitations) become involved in the transport.
The long relaxation time reflects the slow transition rates to, or
from, these states near the energy threshold. For strong
interactions ($g=0.5$ in the figure) clear maxima are seen both at
thresholds for additional plasmon states ($eV = (2g)E_C$) and for
new charge states ($eV = 2E_C$),while for weaker interactions, the
structure at plasmon excitations is almost washed away. This is
due to the fact that for weak interactions, the transition rates
are roughly given by step functions, in contrast to the strong
interaction case when they increase like a power law with a larger
exponent.

\section{Conclusions}
\label{noise-llj::conclusions}

We have investigated the shot noise of a
Luttinger-liquid double-barrier structure.  The non-equilibrium plasmons
on the central segment affect strongly current fluctuations even at low
bias voltages $2\veps_p \leq eV < 2E_C$: The shot noise is enhanced
beyond the Poissonian limit due to the emergence of competing
transport processes, the enhancement
is more pronounced for stronger interactions,
and persists to unexpectedly high frequencies.

\ack
We acknowledge valuable comments from E. V. Sukhorukov and I. V.  Krive.
This work has been supported by the Swedish Foundation for Strategic
Research through the CARAMEL consortium, STINT, the SKORE-A Program, the
eSSC at Postech, and a Korea University Grant. J.U.K.  acknowledges the
hospitality of Department of Physics, Korea University, during his
visit.


\section*{References}


\end{document}